\documentclass[12pt,a4paper]{article}

\usepackage{fullpage}

\usepackage[centertags]{amsmath}
\allowdisplaybreaks[1]

\usepackage{amsbsy}
\usepackage{amsfonts}
\usepackage{amssymb}

\usepackage[dvips]{graphicx}

\usepackage{url}
\usepackage[dvips,hyperindex]{hyperref}
\usepackage{cite}

\usepackage{float}

\begin{document}

\begin{flushright}
% \textsf{\today}
\textsf{9 November 2006}
\\
\textsf{hep-ph/0611125}
\end{flushright}

\vspace{1cm}

\begin{center}
\large
\textbf{Theory and Phenomenology of Neutrino Mixing}
\normalsize
\\[0.5cm]
\large
Carlo Giunti
\normalsize
\\[0.5cm]
INFN, Sezione di Torino, and Dipartimento di Fisica Teorica,
\\
Universit\`a di Torino,
Via P. Giuria 1, I--10125 Torino, Italy
\\[0.5cm]
\begin{minipage}[t]{0.8\textwidth}
\begin{center}
\textbf{Abstract}
\end{center}
We review some fundamental aspects of the theory of neutrino masses and mixing.
The results of neutrino oscillation experiments are
interpreted as evidence of three-neutrino mixing.
Implications for the mixing parameters and the neutrino masses
are discussed, with emphasis on the connection with
the measurements of the absolute values of neutrino masses
in beta decay and neutrinoless double-beta decay experiments
and cosmological observations.
\end{minipage}
\\[0.5cm]
\begin{minipage}[t]{0.8\textwidth}
\footnotesize\tt
\begin{center}
Talks presented at
\\
Tau06, 9th International Workshop on Tau Physics,\\
19-22 September 2006, Pisa, Italy
\\
and
\\
HQL06, International Conference 'Heavy Quarks \& Leptons',\\
16-20 October 2006, Munich, Germany
\end{center}
\end{minipage}
\end{center}

\section{Introduction to Neutrino Masses}
\label{n039}

In the Standard Model (SM) neutrinos are massless.
This is due to the fact that, in the
SM, neutrinos are described by the left-handed chiral fields
$\nu_{eL}$,
$\nu_{{\mu}L}$,
$\nu_{{\tau}L}$
only.
Since the corresponding right-handed fields
$\nu_{eR}$,
$\nu_{{\mu}R}$,
$\nu_{{\tau}R}$
do not exist in the SM,
a Dirac mass term,
\begin{equation}
\mathcal{L}^{\text{D}}
=
\sum_{\alpha,\beta=e,\mu,\tau}
\overline{\nu_{{\alpha}L}} \, M^{\text{D}}_{\alpha\beta} \, \nu_{{\beta}R}
+
\text{H.c.}
\,,
\label{001}
\end{equation}
is precluded.
Here $M^{\text{D}}$ is a complex $3\times3$ mass matrix
(see Refs.~\cite{Bilenky:1987ty,hep-ph/9812360,hep-ph/0211462,hep-ph/0310238}).
On the other hand, the other elementary fermions
(quarks and charged leptons)
are described by left-handed and right-handed chiral fields,
which allow them to have Dirac-type masses
after the spontaneous electroweak symmetry breaking
$ \text{SU(2)}_{L} \times \text{U(1)}_{Y} \to \text{U(1)}_{Q} $ 
generated by the Higgs mechanism.
Note that the off-diagonal terms in the
Dirac mass matrix $M^{\text{D}}$
violate the conservation of the lepton numbers
$\text{L}_{e}$,
$\text{L}_{\mu}$ and
$\text{L}_{\tau}$,
whereas the total lepton number
$ \text{L} = \text{L}_{e}+\text{L}_{\mu}+\text{L}_{\tau} $
is conserved.

In 1937 Ettore Majorana
\cite{Majorana:1937vz}
discovered that a massive neutral fermion can be described by a two-component spinor,
which is simpler than a four-component Dirac spinor.
The fundamental difference of a Majorana fermion with respect to a Dirac fermion
is that for a Majorana fermion the particle and antiparticle states coincide.
In other words,
charge conjugation does not have any effect on a Majorana fermion field.
Since charge conjugation inverts the chirality,
in the SM there are three right-handed neutrino fields
$ (\nu_{{\alpha}L})^{C} \equiv \nu^{C}_{{\alpha}R} $,
for $\alpha=e,\mu,\tau$.
In the Majorana theory,
the right-handed neutrino fields in Eq.~(\ref{001})
are identified with the corresponding charge-conjugated right-handed neutrino fields
$ \nu^{C}_{eR} $,
$ \nu^{C}_{{\mu}R} $,
$ \nu^{C}_{{\tau}R} $,
leading to the Majorana mass term\footnote{
The additional factor $1/2$ is put by hand in order to avoid double counting
in the derivation of the
field equations using the canonical Euler-Lagrange prescription.
}
\begin{equation}
\mathcal{L}_{L}^{\text{M}}
=
\frac{1}{2}
\sum_{\alpha,\beta=e,\mu,\tau}
\overline{\nu_{{\alpha}L}} \, (M_{L}^{\text{M}})_{\alpha\beta} \, \nu^{C}_{{\beta}R}
+
\text{H.c.}
\,,
\label{002}
\end{equation}
with a complex symmetric $3\times3$ mass matrix $M_{L}^{\text{M}}$
(see Refs.~\cite{Bilenky:1987ty,hep-ph/9812360,hep-ph/0211462,hep-ph/0310238}).
Although allowed by the field content of the SM,
this Majorana mass term is forbidden by the gauge symmetries of the SM.
It could be generated by the Vacuum Expectation Value (VEV)
of a Higgs triplet, which is absent in the SM.
Note that the Majorana mass term $\mathcal{L}_{L}^{\text{M}}$
violates the conservation of the total lepton number $\text{L}$
by two units.

Summarizing,
the field content and the gauge symmetries of the SM
hinder the existence of the Dirac and Majorana mass terms in Eqs.~(\ref{001})
and (\ref{002}).
This prediction of the SM is in contradiction with the experimental evidence
of neutrino oscillations,
which are due to neutrino masses and mixing
(see
Refs.~\cite{Bilenky:1978nj,Bilenky:1987ty,hep-ph/9812360,hep-ph/0202058,hep-ph/0211462,hep-ph/0310238,hep-ph/0405172,hep-ph/0506083,hep-ph/0606054}).
Therefore,
it is necessary to extend the SM in order to describe the real world.

The simplest extension of the SM
consists in the introduction of the three right-handed neutrino fields
$\nu_{eR}$,
$\nu_{{\mu}R}$,
$\nu_{{\tau}R}$,
which are singlets under the gauge symmetries of the SM.
In this way,
the neutrino fields become similar to the other elementary fermion fields,
which have both left-handed and right-handed components.
The Dirac mass term in Eq.~(\ref{001})
can be generated by the same Higgs mechanism which generates the Dirac masses
of charged leptons and quarks.
However,
a surprise arises:
the Majorana mass term for the right-handed neutrino fields,
\begin{equation}
\mathcal{L}_{R}^{\text{M}}
=
\frac{1}{2}
\sum_{\alpha,\beta=e,\mu,\tau}
\overline{\nu^{C}_{{\alpha}L}} \, (M_{R}^{\text{M}})_{\alpha\beta} \, \nu_{{\beta}R}
+
\text{H.c.}
\,,
\label{003}
\end{equation}
is invariant under the gauge symmetries of the SM and, hence, allowed.
Therefore,
the seemingly innocuous introduction of right-handed neutrino fields
leads to fundamental new physics: Majorana neutrino masses
and the existence of processes with $|\Delta\text{L}|=2$.

In general,
in a model with
left-handed and right-handed neutrino fields,
the neutrino mass term is of the Dirac-Majorana type
$
\mathcal{L}^{\text{D+M}}
=
\mathcal{L}^{\text{D}}
+
\mathcal{L}_{R}^{\text{M}}
$,
which can be written as
\begin{equation}
\mathcal{L}^{\text{D+M}}
=
\frac{1}{2}
\begin{pmatrix}
\overline{\nu_{L}}
&
\overline{\nu^{C}_{L}}
\end{pmatrix}
\begin{pmatrix}
0 & M^{\text{D}}
\\
(M^{\text{D}})^{T} & M_{R}^{\text{M}}
\end{pmatrix}
\begin{pmatrix}
\nu^{C}_{R}
\\
\nu_{R}
\end{pmatrix}
% \begin{pmatrix}
% \overline{\nu_{L}}
% &
% \overline{\nu^{C}_{L}}
% \end{pmatrix}
% \begin{pmatrix}
% 0 & M^{\text{D}}
% \\
% (M^{\text{D}})^{T} & M_{R}^{\text{M}}
% \end{pmatrix}
% \begin{pmatrix}
% \nu^{C}_{R}
% \\
% \nu_{R}
% \end{pmatrix}
+
\text{H.c.}
\,,
\label{0041}
\end{equation}
where
$
\nu_{L}^{T}
=
\begin{pmatrix}
\nu_{eL}^{T}
&
\nu_{{\mu}L}^{T}
&
\nu_{{\tau}L}^{T}
\end{pmatrix}
$
and
$
\nu_{R}^{T}
=
\begin{pmatrix}
\nu_{eR}^{T}
&
\nu_{{\mu}R}^{T}
&
\nu_{{\tau}R}^{T}
\end{pmatrix}
$.
In the mass matrix,
the $3\times3$ block
which would correspond to $M_{L}^{\text{M}}$
is set to zero because $\mathcal{L}_{L}^{\text{M}}$
is forbidden by the gauge symmetries of the SM,
as explained above.
Since the Dirac mass matrix $M^{\text{D}}$
is generated by the Higgs mechanism of the SM,
its elements are proportional to the VEV of the Higgs doublet,
$ v_{\text{SM}} = \left( \sqrt{2} G_{\text{F}} \right)^{-1/2} = 246 \, \text{GeV} $,
where $G_{\text{F}}$ is the Fermi constant.
Hence,
the elements of $M^{\text{D}}$ are expected to be at most of the order of
$10^{2}\,\text{GeV}$.
This constraint is expressed by saying that they are
``protected'' by the gauge symmetries of the SM.
On the other hand,
since the Majorana mass term of the right-handed neutrino fields
is invariant under the gauge symmetries of the SM,
the elements of $M_{R}^{\text{M}}$
are not protected by the SM gauge symmetries.
In other words, from the SM point of view,
the elements of $M_{R}^{\text{M}}$
could have arbitrarily large values.
If $M_{R}^{\text{M}}$ is generated by the Higgs mechanism at a
high-energy scale of new physics beyond the SM,
the elements of $M_{R}^{\text{M}}$
are expected to be of the order of such new high-energy scale,
which
could be as high as a grand-unification scale of about $ 10^{15} \, \text{GeV} $.
In this case, the total mass matrix can be approximately diagonalized
by blocks,
leading to
a light $3\times3$ mass matrix
\begin{equation}
M_{\text{light}}
\simeq
M^{\text{D}} \, ( M_{R}^{\text{M}} )^{-1} \, {M^{\text{D}}}^{T}
\,,
\label{f393}
\end{equation}
and a heavy $3\times3$ mass matrix
$
M_{\text{heavy}}
\simeq
M_{R}^{\text{M}}
$.
The three light and the three heavy masses
are given,
respectively,
by the eigenvalues of $M_{\text{light}}$ and $M_{\text{heavy}}$.
Therefore, there are three heavy neutrinos which are practically decoupled from
the low-energy physics in our reach
and three light neutrinos
whose masses are suppressed with respect to the elements of the Dirac mass matrix $M^{\text{D}}$
by the small matrix factor $ ( M_{R}^{\text{M}} )^{-1} \, {M^{\text{D}}}^{T} $.
This is the famous see-saw mechanism
\cite{Minkowski:1977sc,Yanagida-SeeSaw-1979,GellMann-Ramond-Slansky-SeeSaw-1979,Mohapatra:1980ia},
which explains naturally the smallness of the three light neutrino masses.
It is important to note that
the see-saw mechanism predicts that massive neutrinos are Majorana particles,
leading to the existence of new measurable phenomena with $|\Delta\text{L}|=2$.
The most accessible is neutrinoless double-$\beta$ decay
(see section~\ref{n044}).

The see-saw mechanism is a particular case
(see Ref.~\cite{hep-ph/9905536})
of the following general argument
\cite{Weinberg:1979sa}
in favor of Majorana massive neutrinos
as a general consequence of new physics beyond the SM
at a high-energy scale $\Lambda$.
The most general effective low-energy Lagrangian can be written as
\begin{equation}
\mathcal{L}_{\text{eff}}
=
\mathcal{L}_{\text{SM}}
+
\frac{\Omega_{5}}{\Lambda}
+
\frac{\Omega_{6}}{\Lambda^2}
+
\ldots
\,,
\label{021}
\end{equation}
where $\mathcal{L}_{\text{SM}}$ is the SM Lagrangian.
The additional non-SM terms contain the field operators
$ \Omega_{5} $,
$ \Omega_{6} $,
$ \ldots $,
which have energy dimension larger than four, as indicated by the index.
These operators contain SM fields only.
Furthermore,
they are constrained to be invariant under the SM gauge symmetries,
because the new high-energy theory by which they are generated
is an extension of the SM.
They are not included in $\mathcal{L}_{\text{SM}}$,
because they are not renormalizable
(similarly to the Fermi Lagrangian,
which is the effective non-renormalizable Lagrangian of weak interactions
for energies much smaller than $v_{\text{SM}}$).
Since each Lagrangian term must have energy dimension equal to four,
the non-SM  terms in Eq.~(\ref{021})
are suppressed by appropriate negative powers of the high-energy scale $\Lambda$.
The less-suppressed non-SM term is the 5-D operator
\begin{equation}
\Omega_{5}
=
\sum_{\alpha\beta}
g_{\alpha\beta}
\,
( L^{T}_{\alpha L} \, \sigma_{2} \, \Phi )
\, \mathcal{C}^{\dagger} \,
( \Phi^{T} \, \sigma_{2} \, L_{\beta L} )
+
\text{H.c.}
\,,
\label{022}
\end{equation}
where $L_{\alpha L}$, $\Phi$, $\mathcal{C}$ and $\sigma_{i}$ are,
respectively,
the left-handed lepton doublets ($\alpha=e,\mu,\tau$),
the Higgs doublet,
the charge-conjugation matrix
and the Pauli matrices ($i=1,2,3$).
At the electroweak symmetry breaking,
$\Omega_{5}$ generates a Majorana mass term of the type in Eq.~(\ref{002}),
with the mass matrix
\begin{equation}
(M_{L}^{\text{M}})_{\alpha\beta}
=
\frac{v_{\text{SM}}^{2}}{\Lambda} \, g_{\alpha\beta}
\ll
v_{\text{SM}}
=
246 \, \text{GeV}
\,.
\label{023}
\end{equation}
Hence,
the neutrino masses
are naturally suppressed by the very small ratio $v_{\text{SM}}/\Lambda$
with respect to
the masses of the charged leptons and quarks, which are proportional to $v_{\text{SM}}$.
It is remarkable that
the 5-D operator in Eq.~(\ref{022})
is unique,
in contrast to the multiplicity of 6-D operators
(see Ref.~\cite{Costa:1986vk}),
which include operators for nucleon decay.
Hence,
Majorana neutrino masses provide the most accessible
window on new physics beyond the SM.

\begin{figure}[t!]
\begin{center}
\renewcommand{\arraystretch}{2}
\setlength{\tabcolsep}{2cm}
\begin{tabular}{cc}
\begin{minipage}[c]{0.25\textwidth}
\includegraphics*[bb=175 469 415 779, width=\linewidth]{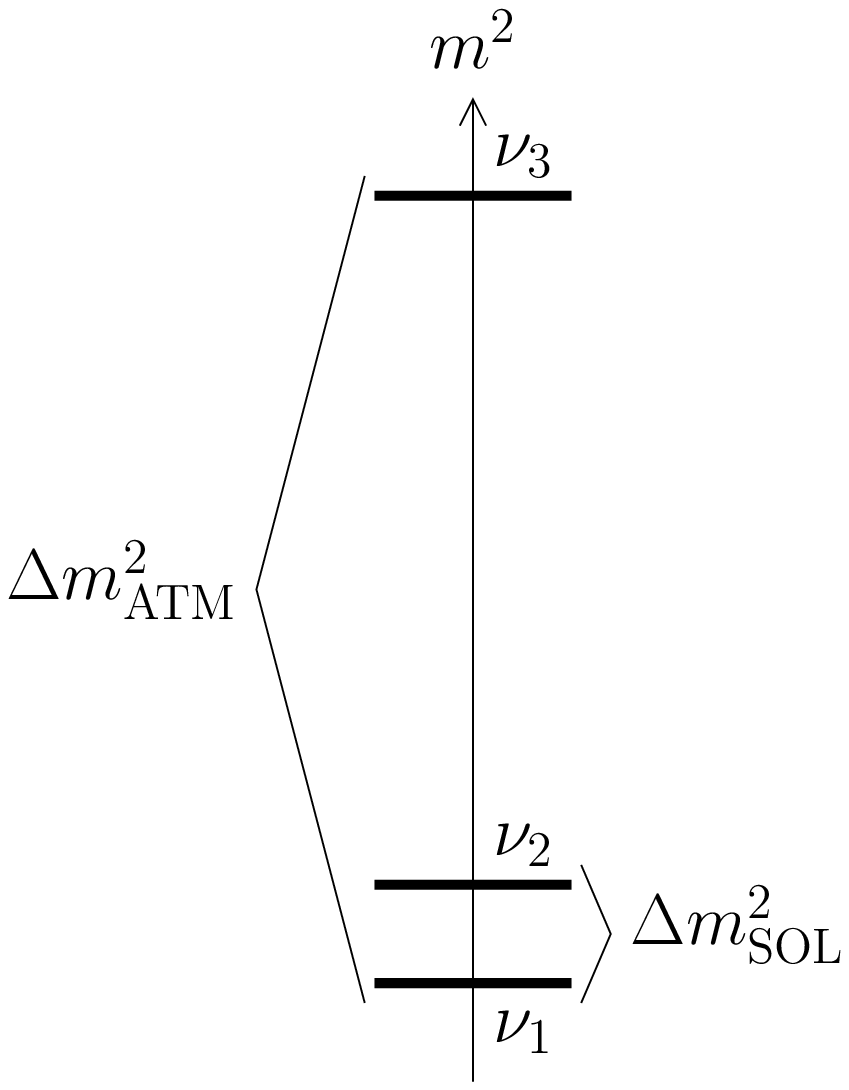}
\end{minipage}
&
\begin{minipage}[c]{0.25\textwidth}
\includegraphics*[bb=180 469 420 779, width=\linewidth]{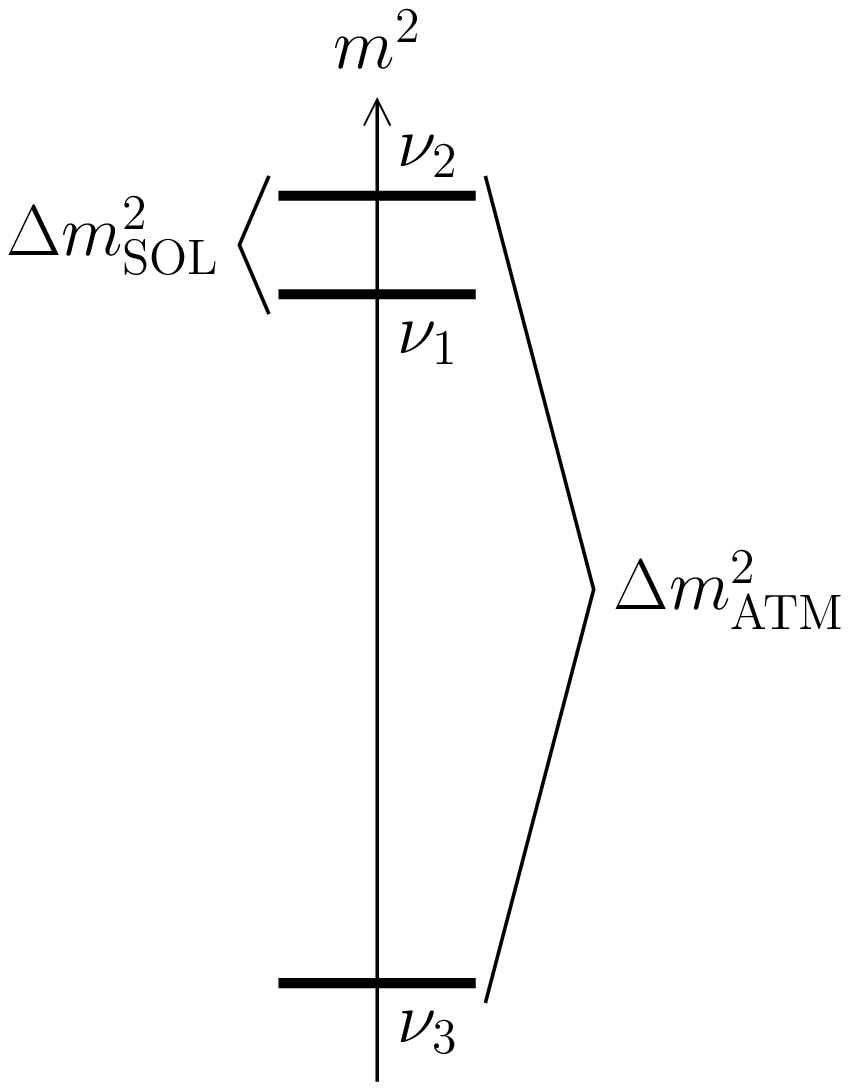}
\end{minipage}
\\
\underline{NORMAL}
&
\underline{INVERTED}
\end{tabular}
\end{center}
\caption{ \label{m008}
The two three-neutrino schemes allowed by the hierarchy
$\Delta{m}^{2}_{\text{SOL}} \ll \Delta{m}^{2}_{\text{ATM}}$.
}
\end{figure}

\section{Three-Neutrino Mixing}
\label{n040}

The mass matrix of the three light neutrinos
(either Dirac or Majorana)
can be diagonalized through the unitary transformation
\begin{equation}
\nu_{{\alpha}L}
=
\sum_{k=1}^{3} U_{{\alpha}k} \, \nu_{kL}
\,,
\label{101}
\end{equation}
where $U$ is the $3\times3$ unitary mixing matrix.
An important consequence of neutrino mixing
is the existence of neutrino flavor oscillations,
which depend on the elements of the mixing matrix
and on the squared-mass differences
$ \Delta{m}^{2}_{kj} \equiv m_{k}^{2} - m_{j}^{2} $.
Neutrino oscillations have been observed
(see
Refs.~\cite{hep-ph/9812360,hep-ph/0202058,hep-ph/0211462,hep-ph/0310238,hep-ph/0405172,hep-ph/0506083,hep-ph/0606054})
in solar and reactor neutrino experiments
($\nu_{e}\to\nu_{\mu,\tau}$),
with a squared-mass difference
\cite{hep-ph/0506083}
\begin{equation}
\Delta{m}^{2}_{\text{SOL}}
=
7.92 \left( 1 \pm 0.09 \right) \times 10^{-5} \, \text{eV}^2
\quad
[2\sigma]
\,,
\label{SOL}
\end{equation}
and
in atmospheric and accelerator neutrino experiments
($\nu_{\mu}\to\nu_{\tau}$),
with a squared-mass difference
\cite{hep-ph/0608060}
\begin{equation}
\Delta{m}^{2}_{\text{ATM}}
=
2.6 \left( 1 {}^{+0.14}_{-0.15} \right) \times 10^{-3} \, \text{eV}^2
\quad
[2\sigma]
\,.
\label{ATM}
\end{equation}
Hence, there is a hierarchy of squared-mass differences:
\begin{equation}
\Delta m^2_{\text{ATM}}
\simeq
30 \, \Delta m^2_{\text{SOL}}
\,.
\label{102}
\end{equation}
This hierarchy is easily accommodated in the framework of three-neutrino mixing,
in which there are two independent squared-mass differences.
We label the neutrino masses in order to have
\begin{align}
\null & \null
\Delta{m}^2_{\text{SOL}}
=
\Delta{m}^2_{21}
\,,
\label{103}
\\
\null & \null
\Delta{m}^2_{\text{ATM}}
\simeq
|\Delta{m}^2_{31}|
\simeq
|\Delta{m}^2_{32}|
\,.
\label{104}
\end{align}
The two possible schemes are illustrated in Fig.~\ref{m008}.
They differ by the sign of
$
\Delta{m}^2_{31}
\simeq
\Delta{m}^2_{32}
$.

\begin{figure}[t!]
\begin{center}
\includegraphics*[bb=227 658 364 770, width=0.3\textwidth]{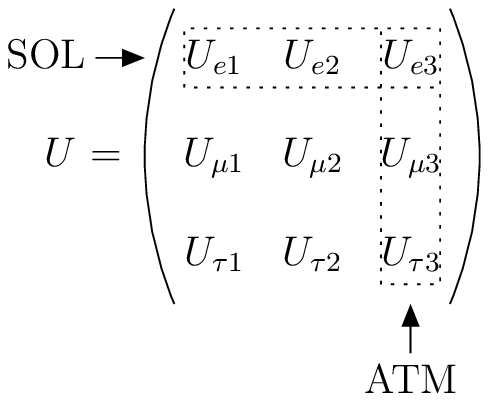}
\end{center}
\caption{ \label{m135}
Schematic description of the contributions of the elements of the mixing matrix to solar (SOL) and atmospheric (ATM)
neutrino oscillations.
}
\end{figure}

Information on neutrino mixing
is traditionally obtained from the analysis of the experimental
data in the framework of an effective
two-neutrino mixing scheme,
in which oscillations depend on only one squared-mass difference
($\Delta{m}^2$)
and one mixing angle
($\vartheta$).
This approximation is allowed
\cite{hep-ph/9802201}
by the smallness of
$|U_{e3}|$,
which is the only element of the mixing matrix which affects
both the solar-reactor and atmospheric-accelerator
oscillations, as illustrated in Fig.~\ref{m135}.
In fact,
solar and reactor experiments
have observed the disappearance of electron neutrinos,
which depends
only on the elements of the
mixing matrix which connect $\nu_{e}$ with the three massive neutrinos:
$U_{e1}$,
$U_{e2}$ and
$U_{e3}$.
On the other hand, the hierarchy of squared-mass differences
in Eq.~(\ref{102})
implies that $\nu_{1}$ and $\nu_{2}$
are practically the same in atmospheric and accelerator
oscillations and contribute through
$ |U_{\alpha1}|^{2} + |U_{\alpha2}|^{2} = 1 - |U_{\alpha3}|^{2} $.
Hence these oscillations depend only on the third column of
the elements of the mixing matrix.

The smallness of $|U_{e3}|$ is known from the
results of the CHOOZ and Palo Verde
experiments
(see Ref.~\cite{hep-ph/0107277}),
leading to
\cite{hep-ph/0608060}
\begin{equation}
|U_{e3}|^{2} = 0.008 {}^{+0.023}_{-0.008}
\qquad
[2\sigma]
\,.
\label{m140}
\end{equation}
In this case,
in the standard parameterization of
the mixing matrix
(see Ref.~\cite{hep-ph/0310238}),
we have
$ |U_{e3}|^{2} = \sin^2 \vartheta_{13} $,
the effective mixing angle measured in
solar and reactor experiment
is approximately equal to $\vartheta_{12}$
and the effective mixing angle measured in
atmospheric and accelerator experiment
is approximately equal to $\vartheta_{23}$.
An analysis of the data yields large values for
$\vartheta_{12}$ and $\vartheta_{23}$
\cite{hep-ph/0506083,hep-ph/0608060}:
\begin{align}
\null & \null
\sin^2\vartheta_{12}
=
0.314 \left( 1 {}^{+0.18}_{-0.15} \right)
\qquad
[2\sigma]
\,,
\label{151}
\\
\null & \null
\sin^2\vartheta_{23}
=
0.45 \left( 1 {}^{+0.35}_{-0.20} \right)
\qquad
[2\sigma]
\,.
\label{152}
\end{align}
The mixing angle $\vartheta_{23}$ is close to maximal ($\pi/4$).
The mixing angle $\vartheta_{12}$ is large,
but less than maximal.

From the determination of the mixing angles,
it is possible to reconstruct the allowed ranges for the
elements of the mixing matrix:
at $2\sigma$ we have
\begin{equation}
|U|_{2\sigma}
\simeq
\begin{pmatrix}
0.78-0.86 & 0.51-0.61 & 0.00-0.18 \\
0.21-0.57 & 0.41-0.74 & 0.59-0.78 \\
0.19-0.56 & 0.39-0.72 & 0.62-0.80 \\

\end{pmatrix}
\,.
\label{153}
\end{equation}
One can see that all the elements of the mixing matrix are large,
except $|U_{e3}|$,
for which we have only an upper bound.

A mixing matrix of the type in Eq.~(\ref{153}),
with two large mixing angles
($\vartheta_{12}$ and $\vartheta_{23}$),
is called ``bilarge''.
Several future experiments are
aimed at a measurement of the small mixing angle $\vartheta_{13}$
(see Ref.~\cite{hep-ph/0606111}),
whose finiteness is crucial for the existence of
CP violation in the lepton sector,
for the possibility to measure matter effects with future neutrino beam passing through the Earth
and
for the possibility to distinguish the normal and inverted schemes in
future oscillation experiments.

\begin{figure}[t!]
\begin{minipage}[t]{0.47\textwidth}
\begin{center}
\includegraphics*[bb=102 430 445 754, width=0.95\textwidth]{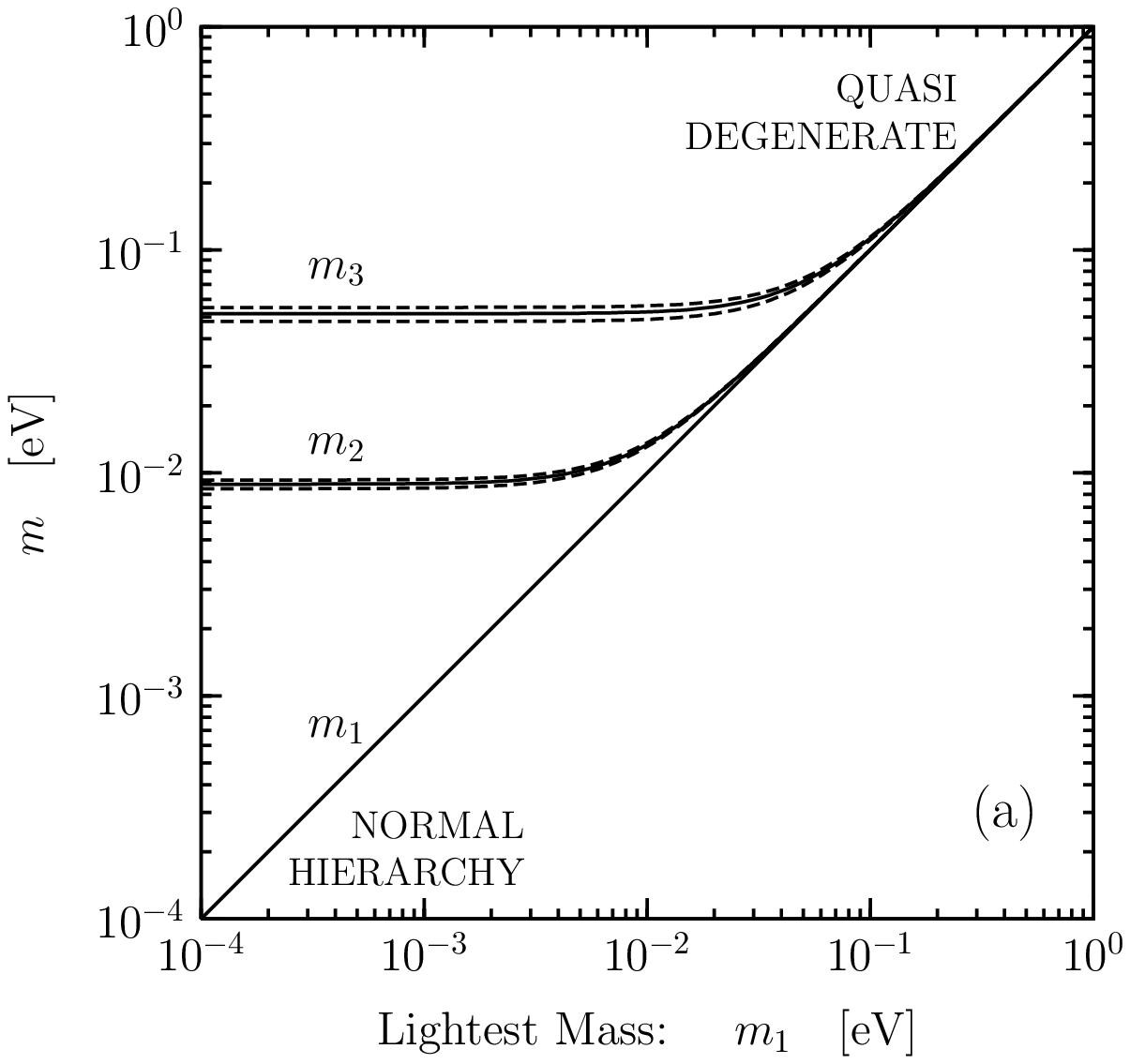}
\end{center}
\end{minipage}
\hfill
\begin{minipage}[t]{0.47\textwidth}
\begin{center}
\includegraphics*[bb=102 430 445 754, width=0.95\textwidth]{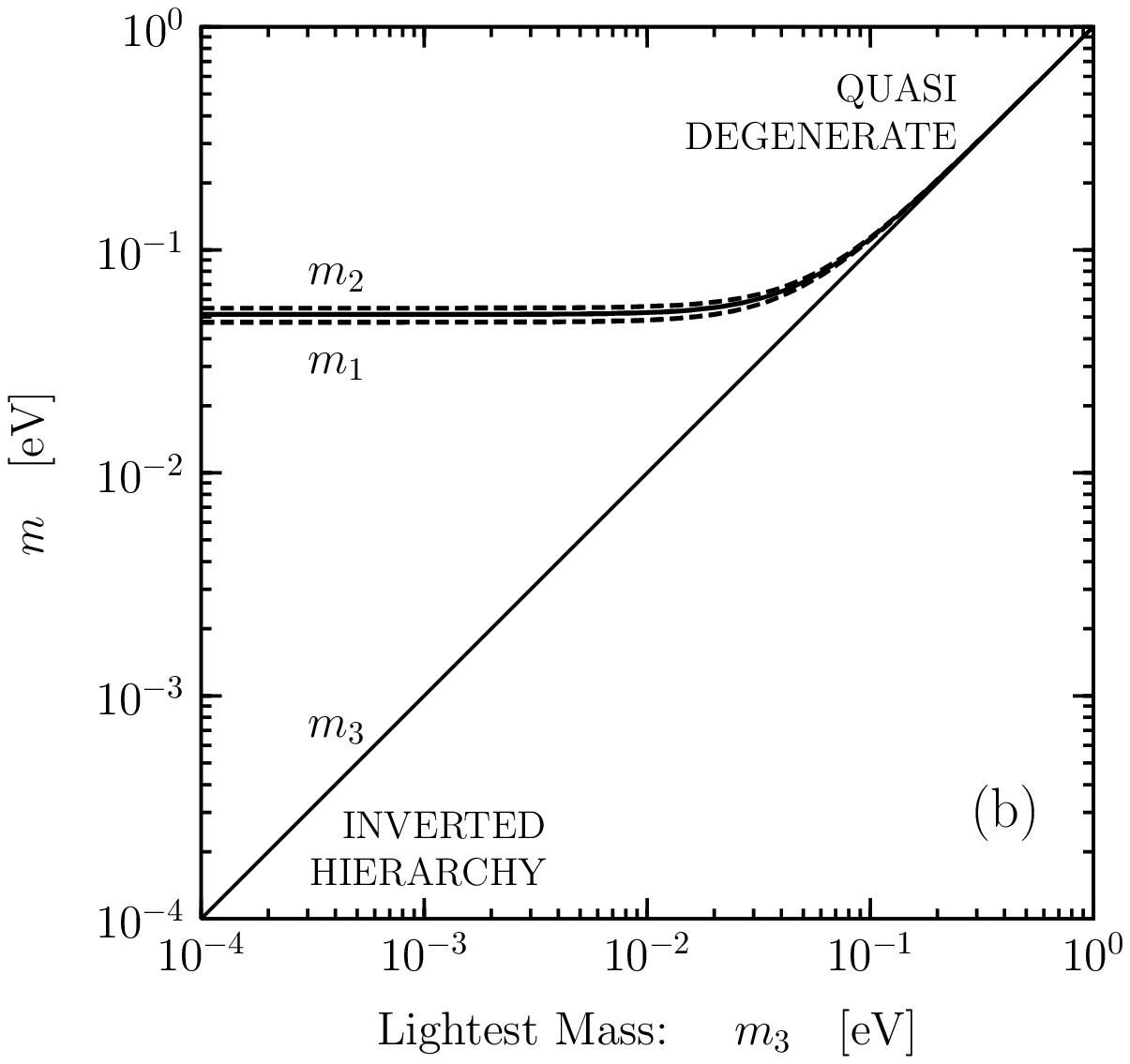}
\end{center}
\end{minipage}
\caption{ \label{3ma}
Values of neutrino masses as functions
of the lightest mass,
$m_{1}$ in the normal scheme (a) and $m_3$ in the inverted scheme (b).
Solid lines correspond to the best-fit.
Dashed lines enclose $2\sigma$ ranges.
}
\end{figure}

As a first approximation,
it is instructive to consider
$\vartheta_{13}=0$.
In this case,
the mixing matrix is given by
\begin{equation}
U
=
\begin{pmatrix}
c_{\vartheta_{12}} & s_{\vartheta_{12}} & 0
\\
- s_{\vartheta_{12}} c_{\vartheta_{23}} & c_{\vartheta_{12}} c_{\vartheta_{23}} & s_{\vartheta_{23}}
\\
s_{\vartheta_{12}} s_{\vartheta_{23}} & - c_{\vartheta_{12}} s_{\vartheta_{23}} & c_{\vartheta_{23}}
\end{pmatrix}
\,,
\label{154}
\end{equation}
where
$c_{\vartheta_{ij}}\equiv\cos\vartheta_{ij}$
and
$s_{\vartheta_{ij}}\equiv\sin\vartheta_{ij}$.
Choosing the attractive values
\begin{equation}
\sin^2\vartheta_{12} = \frac{1}{3}
\,,
\qquad
\sin^2\vartheta_{23} = \frac{1}{2}
\,,
\label{155}
\end{equation}
which are within the ranges in Eqs.~(\ref{151}) and (\ref{152}),
we have the so-called ``tri-bimaximal``
mixing matrix
\cite{hep-ph/0202074}
\begin{equation}
U
=
\begin{pmatrix}
\sqrt{2/3} & 1/\sqrt{3} & 0
\\
- 1/\sqrt{6} & 1/\sqrt{3} & 1/\sqrt{2}
\\
1/\sqrt{6} & - 1/\sqrt{3} & 1/\sqrt{2}
\end{pmatrix}
\,.
\label{156}
\end{equation}
The name is due to the fact that the magnitudes of all the elements of the second column are equal
(trimaximal mixing)
and the third column have only two finite elements, which have the same magnitude
(bimaximal mixing).

In the approximation in Eq.~(\ref{154}),
we have
\begin{equation}
\nu_e = c_{\vartheta_{12}} \nu_{1} + s_{\vartheta_{12}} \nu_{2}
\,,
\label{157}
\end{equation}
which is a two-neutrino mixing relation.
In oscillations,
electron neutrinos can transform in the orthogonal state
\begin{equation}
\nu_{\perp}
=
- s_{\vartheta_{12}} \nu_{1} + c_{\vartheta_{12}} \nu_{2}
=
c_{\vartheta_{23}} \nu_\mu - s_{\vartheta_{23}} \nu_\tau
\,.
\label{158}
\end{equation}
Hence,
The state in which solar and reactor electron neutrinos transform
is a superposition of $\nu_\mu$ and $\nu_\tau$
determined by the atmospheric mixing angle $\vartheta_{23}$.
The closeness of $\vartheta_{23}$ to maximal mixing
implies an approximate equal amount of $\nu_\mu$ and $\nu_\tau$.
If one further takes into account that the SNO experiment
measured a suppression of about $1/3$
of the solar $\nu_{e}$ flux for $ E \gtrsim 6 \, \text{MeV} $,
it follows that the flux of high-energy solar neutrinos on the Earth is composed of
an approximately equal amount of $\nu_{e}$, $\nu_\mu$ and $\nu_\tau$.

\section{The Absolute Scale of Neutrino Masses}
\label{n041}

Since neutrino oscillations depend on the differences of the
squared neutrino masses,
other types of experiments are needed
in order to determine
the absolute values of neutrino masses.
However,
what is really unknown from the results of neutrino oscillation experiments
is only one mass,
since the other masses can be determined from the known difference of the
squared neutrino masses.
In the three-neutrino mixing schemes in Fig.~\ref{m008},
it is convenient to choose as unknown the lightest mass
($m_{1}$ in the normal scheme and $m_{3}$ in the inverted scheme)
and plot the masses as shown in Fig.~\ref{3ma}.
One can see that, in the normal scheme,
if $m_{1}$ is small, there is a normal mass hierarchy
$ m_{1} \ll m_{2} \ll m_{3} $.
On the other hand,
in the inverted scheme, if $m_{3}$ is small, there is a so-called ``inverted mass hierarchy''
$ m_{3} \ll m_{1} \lesssim m_{2} $,
since $m_{1}$ and $m_{2}$ are separated by the small solar mass splitting.
In both schemes,
the three neutrino masses are quasi-degenerate for
\begin{equation}
m_{3} \gtrsim m_{2} \gtrsim m_{1}
\gg
\sqrt{\Delta{m}^2_{\text{ATM}}} \simeq 5 \times 10^{-2} \, \text{eV}
\,.
\label{161}
\end{equation}

In the next three subsections,
we discuss the tree most efficient methods for the determination
of the absolute scale of neutrino masses:
$\beta$ decay, cosmological observations and
neutrinoless double-$\beta$ decay.

\subsection{Beta Decay}
\label{n042}

The measurement of the energy spectrum of electrons emitted
in nuclear $\beta$ decay provides a robust kinematical measurement
of the effective electron neutrino mass.

\begin{figure}[t!]
\begin{center}
\includegraphics*[bb=113 544 470 737, width=0.7\textwidth]{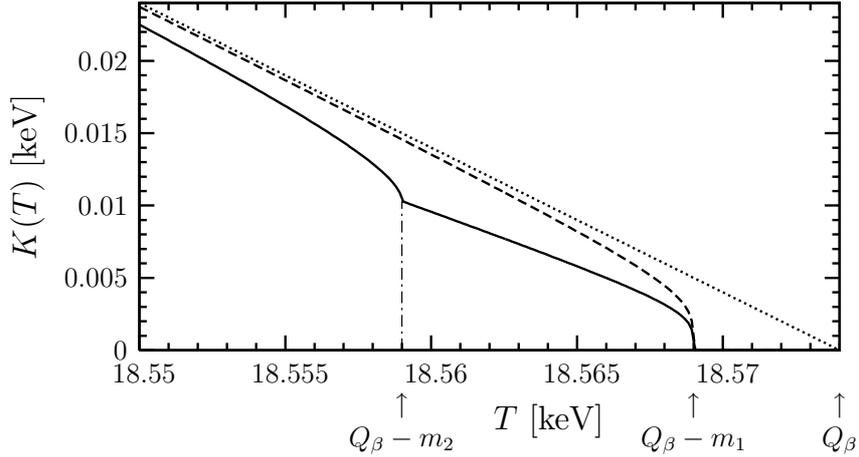}
\end{center}
\caption{ \label{n024}
Kurie plot for tritium $\beta$ decay.
Dotted line: the linear Kurie function for $m_{\nu_{e}}=0$.
Dashed line: Kurie function in Eq.~(\ref{n019}) for $m_{\nu_{e}} = 5 \, \text{eV}$.
Solid line: Kurie function in Eq.~(\ref{n029}) for two-neutrino mixing
with $m_{1} = 5 \, \text{eV}$, $m_{2} = 15 \, \text{eV}$ and $\vartheta=\pi/4$.
}
\end{figure}

Let us consider first, for simplicity,
a massive electron neutrino without mixing.
In this case,
the differential decay rate in allowed\footnote{
Allowed $\beta$-decays
are characterized by the independence of the nuclear matrix element
from the electron energy.
}
$\beta$-decays
is proportional to the square of the Kurie function
\begin{equation}
K(T)
=
\left[
\left( Q - T \right)
\sqrt{ \left( Q - T \right)^2 - m_{\nu_{e}}^{2} }
\right]^{1/2}
\,,
\label{n019}
\end{equation}
where
$
Q
=
M_{i}
-
M_{f}
-
m_{e}
$
($M_{i}$ and $M_{f}$ are, respectively,
the masses of the initial and final nuclei
and $m_{e}$ is the electron mass)
and
$ T = E_{e} - m_{e} $
is the electron kinetic energy.
If $m_{\nu_{e}}=0$,
the Kurie function is a decreasing linear function of $T$,
going to zero at the so-called ``end-point'' of the spectrum, $T=Q$,
as illustrated by the dotted line in Fig.~\ref{n024} for tritium $\beta$ decay.
A small electron neutrino mass affects the electron spectrum near the end-point,
which shifts to $T=Q-m_{\nu_{e}}$,
as shown by the dashed line in Fig.~\ref{n024}.
Therefore,
in practice, information on the value of the neutrino mass
is obtained looking for a distortion of the Kurie plot
with respect to the linear function
near the end-point.
Using this technique,
the Mainz tritium experiment \cite{hep-ex/0412056}
obtained the most stringent upper bound on the electron neutrino mass:
\begin{equation}
m_{\nu_{e}}
<
2.3 \, \text{eV}
\quad
[95\% \, \text{CL}]
\,.
\label{n021}
\end{equation}
The Troitzk tritium experiment \cite{Lobashev:1999tp}
obtained the comparable bound
$
m_{\nu_{e}}
<
2.5 \, \text{eV}
$
[95\% CL].
The main reason why tritium $\beta$-decay experiments
are the most sensitive to the electron neutrino mass
is that
tritium $\beta$-decay has one of the smallest $Q$-values among all known $\beta$-decays.
Since
the relative number of events occurring in an interval of energy
$\Delta{T}$
below the end-point is proportional to
$ \left( \Delta{T} / Q \right)^{3} $,
a small $Q$-value is desirable
for a maximization of the fraction of decay events that occur
near the end-point of the spectrum.
Moreover, tritium $\beta$-decay is a superallowed
transition between mirror nuclei\footnote{
Superallowed transitions are allowed transitions between
nuclei belonging to the same isospin multiplet.
Mirror nuclei are pairs of nuclei which have equal numbers of protons and neutrons
plus an extra proton in one case and an extra neutron in the other.
In this case, the overlap of the initial and final
nuclear wave functions is close to one,
leading to a large nuclear matrix element.
}
with a relatively short half-life
(about 12.3 years),
which implies an acceptable number of observed events
during the experiment lifetime.
Another advantage of tritium $\beta$-decay is that
the atomic structure is less complicated than those of heavier atoms,
leading to a more accurate calculation of atomic effects.

In the case of neutrino mixing,
the Kurie function is given by
\begin{equation}
K(T)
=
\Big[
\left( Q - T \right)
\sum_{k=1}^{3} |U_{ek}|^2 \, \sqrt{ \left( Q - T \right)^2 - m_{k}^{2} }
\Big]^{1/2}
\,.
\label{n029}
\end{equation}
This is a function of 5 parameters,
the three neutrino masses and two mixing parameters
(the unitarity of the mixing matrix implies that
$ \sum_{k=1}^{3} |U_{ek}|^{2} = 1 $).
The main characteristics of the
distortion of the Kurie function with respect to the linear function corresponding to massless neutrinos are:
\renewcommand{\labelenumi}{(\theenumi)}
\renewcommand{\theenumi}{\alph{enumi}}
\begin{enumerate}
\item
A shift of the end-point of the spectrum from $T=Q$ to
$T=Q-m_{\text{lht}}$,
calling $\nu_{\text{lht}}$ the lightest massive neutrino component
of $\nu_{e}$
(if $U_{e3}=0$, $\nu_{\text{lht}}=\nu_{1}$ in both the normal and inverted schemes;
otherwise,
$\nu_{\text{lht}}=\nu_{1}$ in the normal scheme
and
$\nu_{\text{lht}}=\nu_{3}$ in the inverted scheme).
\item
Kinks at the electron kinetic energies
$T_{k} = Q - m_{k}$,
for $\nu_{k}\neq\nu_{\text{lht}}$,
with corresponding strength determined by the value of
$|U_{ek}|^{2}$.
\end{enumerate}
This behavior of the Kurie function is illustrated by the solid line in Fig.~\ref{n024},
which describes the case of two-neutrino mixing
($U_{e3}=0$)
with $m_{1} = 5 \, \text{eV}$, $m_{2} = 15 \, \text{eV}$ and $\vartheta=\pi/4$
($|U_{e1}|^{2}=|U_{e2}|^{2}=1/2$).

If, in the future, effects of the neutrino masses will be
discovered in tritium or other $\beta$-decay experiments,
a precise analysis of the data may reveal kinks of the Kurie function
due to mixing of the electron neutrino with more than one massive neutrino.
In this case,
the data will have to be analyzed using Eq.~(\ref{n029}).

However,
so far tritium experiments did not find any effect of the neutrino masses
and their data have been analyzed in terms of the one-generation Kurie function in Eq.~(\ref{n019}),
leading to the upper bound in Eq.~(\ref{n021}).
How this result can be interpreted in the framework of three-neutrino mixing,
in which Eq.~(\ref{n029}) holds?
The exact expression of $K(T)$ in Eq.~(\ref{n029})
cannot be reduced to the one-generation Kurie function in Eq.~(\ref{n019}).
In order to achieve such a reduction in an approximate way,
one must note that,
if an experiment does not find any effect of the neutrino masses,
its resolution for the measurement of $ Q_{\beta} - T $
is much larger than the values of the neutrino masses.
Considering $ m_{k} \ll Q_{\beta} - T $,
we have
\begin{align}
K^{2}
=
\null & \null
\left( Q - T \right)^{2}
\sum_{k} |U_{ek}|^{2}
\sqrt{ 1 - \frac{ m_{k}^{2} }{ \left( Q - T \right)^{2} } }
\simeq
\left( Q - T \right)^{2}
\sum_{k} |U_{ek}|^{2}
\left[ 1 - \frac{1}{2} \frac{ m_{k}^{2} }{ \left( Q - T \right)^{2} } \right]
\nonumber
\\
=
\null & \null
\left( Q - T \right)^{2}
\left[ 1 - \frac{1}{2} \frac{ m_{\beta}^{2} }{ \left( Q - T \right)^{2} } \right]
\simeq
\left( Q - T \right)^{2}
\sqrt{ 1 - \frac{ m_{\beta}^{2} }{ \left( Q - T \right)^{2} } }
\nonumber
\\
=
\null & \null
\left( Q - T \right)
\sqrt{ \left( Q - T \right)^{2} - m_{\beta}^{2} }
\,,
\label{n030}
\end{align}
with $m_{\beta}$ given by
\begin{equation}
m_{\beta}^{2} = \sum_{k} |U_{ek}|^{2} m_{k}^{2}
\,.
\label{n031}
\end{equation}
The approximate expression of $K(T)$ in terms of $m_{\beta}$ is the same as
the expression in Eq.~(\ref{n019}) of the one-generation Kurie function in terms of $m_{\nu_{e}}$.
Therefore,
$m_{\beta}$ can be considered as
the effective electron neutrino mass in $\beta$-decay.
In the case of three-neutrino mixing,
the upper bound in Eq.~(\ref{n021})
must be interpreted as a bound on $m_{\beta}$:
\begin{equation}
m_{\beta}
<
2.3 \, \text{eV}
\quad
[95\% \, \text{CL}]
\,.
\label{n032}
\end{equation}
If the future experiments do not find any effect of neutrino masses,
they will provide more stringent bounds on the value of $m_{\beta}$.

\begin{figure}[t!]
\begin{center}
\setlength{\tabcolsep}{0cm}
\begin{tabular*}{\textwidth}{l@{\extracolsep{\fill}}r}
\includegraphics*[bb=102 430 445 754, width=0.49\textwidth]{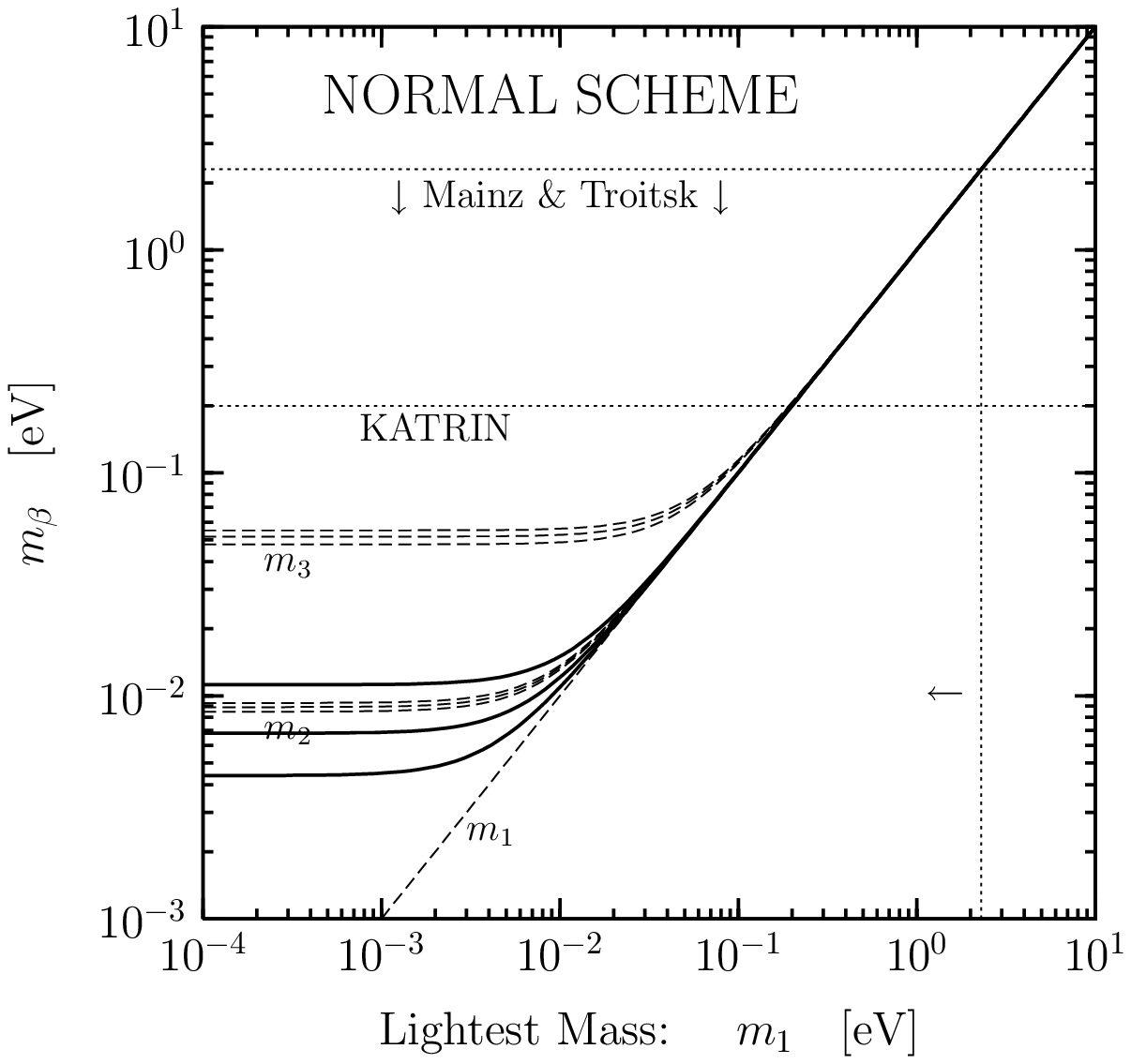}
&
\includegraphics*[bb=102 430 445 754, width=0.49\textwidth]{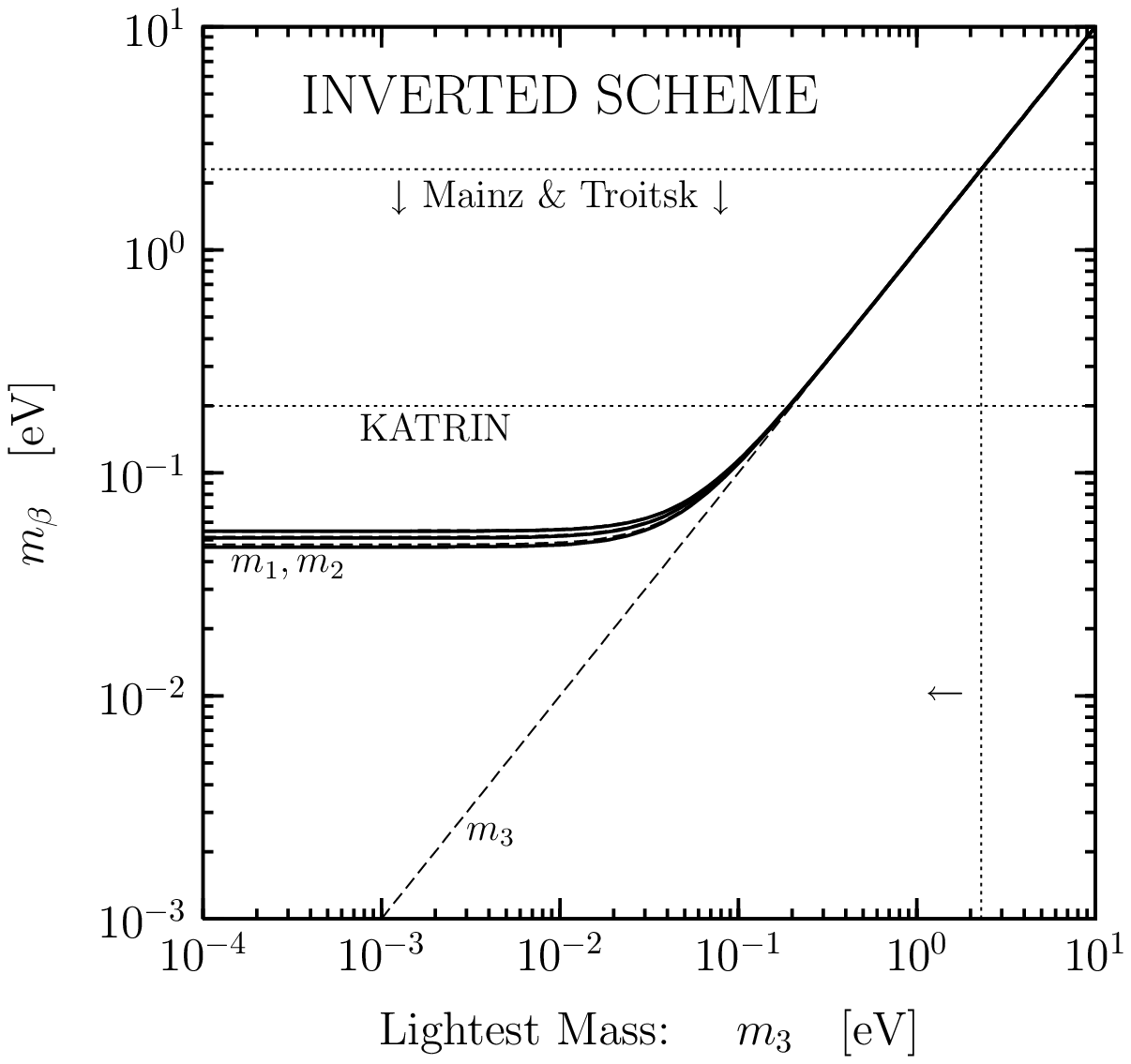}
\end{tabular*}
\end{center}
\caption{ \label{n033}
Effective neutrino mass $m_{\beta}$
in tritium $\beta$-decay experiments as a function
of the lightest mass
($m_{1}$ in the normal scheme and $m_{3}$ in the inverted scheme; see Fig.~\ref{m008}).
Middle solid lines correspond to the best-fit values of the oscillation parameters.
Extreme solid lines enclose $2\sigma$ ranges.
Dashed lines show the best-fit values and $2\sigma$ ranges of individual masses.
In the inverted scheme, the best-fit values and $2\sigma$ ranges of $m_{1}$ and $m_{2}$
are practically the same and coincide with the best-fit value and $2\sigma$ range of $m_{\beta}$.
}
\end{figure}

In the standard parameterization of the mixing matrix, we have
\begin{equation}
m_{\beta}^{2}
=
c_{12}^{2} \, c_{13}^{2} \, m_{1}^{2}
+
s_{12}^{2} \, c_{13}^{2} \, m_{2}^{2}
+
s_{13}^{2} \, m_{3}^{2}
\,.
\label{n034}
\end{equation}
Although neutrino oscillation experiments do not give information
on the absolute values of neutrino masses,
they give information on the squared-mass differences
$\Delta{m}^{2}_{21}$ and
$\Delta{m}^{2}_{31}$
and on the mixing angles $\vartheta_{12}$ and $\vartheta_{13}$
(see Eqs.~(\ref{SOL}), (\ref{ATM}), (\ref{m140}), (\ref{151}) and (\ref{152})).
As shown in Fig.~\ref{3ma},
the values of the neutrino masses can be determined
as functions of the lightest mass
($m_{1}$ in the normal scheme and $m_{3}$ in the inverted scheme).
Therefore,
also $m_{\beta}$ can be considered as a function of the lightest mass,
as shown in Fig.~\ref{n033}.
The middle solid lines correspond to the best fit
and the extreme solid lines delimit the $2\sigma$ allowed range.
We have also shown with dashed lines
the best-fit and $2\sigma$ ranges of the neutrino masses
(same as in Fig.~\ref{3ma}),
which help to understand their contribution to
$m_{\beta}$.

From Fig.~\ref{n033} one can see that, in the case of a normal mass hierarchy
(normal scheme with
$m_{1} \ll m_{2} \ll m_{3}$),
the main contribution to $m_{\beta}$
is due to $m_{2}$ or $m_{3}$ or both,
because the upper limit for
$m_{\beta}$
is larger than the upper limit for
$m_{2}$.
In the case of an inverted mass hierarchy
(inverted scheme with
$m_{3} \ll m_{1} \lesssim m_{2}$),
$m_{\beta}$
has practically the same value as $m_{1}$ and $m_{2}$.
In the case of a quasi-degenerate spectrum,
$m_{\beta}$
coincides with the approximately equal value of the three neutrino masses
in both the normal and inverted schemes.

Figure~\ref{n033}
shows that the present experiments
and the future KATRIN experiment \cite{hep-ex/0309007},
with an expected sensitivity of about $0.2 \, \text{eV}$,
give information on the absolute values of neutrino masses
in the quasi-degenerate region
in both the normal and inverted schemes.
From the Mainz upper bound in Eq.~(\ref{n021}),
for the individual neutrino masses we obtain
\begin{equation}
m_{k}
<
2.3 \, \text{eV}
\quad
[95\% \, \text{CL}]
\,,
\label{n036}
\end{equation}
with $k=1,2,3$.

One can note from Fig.~\ref{n033}
that the allowed ranges of $m_{\beta}$ in the normal and inverted schemes in
the case of a mass hierarchy
are quite different and non overlapping:
the lower limit for $m_{\beta}$ in the inverted scheme is
about $4.7 \times 10^{-2} \, \text{eV}$,
whereas the upper limit for $m_{\beta}$ in the normal scheme is
about $1.1 \times 10^{-2} \, \text{eV}$.
If future experiments find an upper bound for $m_{\beta}$
which is smaller than about
$4.7 \times 10^{-2} \, \text{eV}$,
the inverted scheme will be excluded,
leaving the normal scheme as the only possibility.

Figure~\ref{n033} shows also that $\beta$-decay experiments will not have to improve
indefinitely for finding the effects of neutrino masses:
the ultimate sensitivity is set at about
$4 \times 10^{-3} \, \text{eV}$,
which is the lower bound for $m_{\beta}$ in the case of a normal mass hierarchy.
Of course, when some $\beta$-decay experiment will reveal the effects of neutrino masses,
a more complicated analysis using the expression of $K(T)$ in Eq.~(\ref{n029})
will be needed.
In that case, it may be possible
to distinguish between the normal and inverted schemes
even if both are allowed
(i.e.\ $m_{\beta} \gtrsim 4.7 \times 10^{-2} \, \text{eV}$).

\begin{figure}[t!]
\begin{minipage}[t]{0.47\textwidth}
\begin{center}
\includegraphics*[bb=102 430 437 748, width=0.95\textwidth]{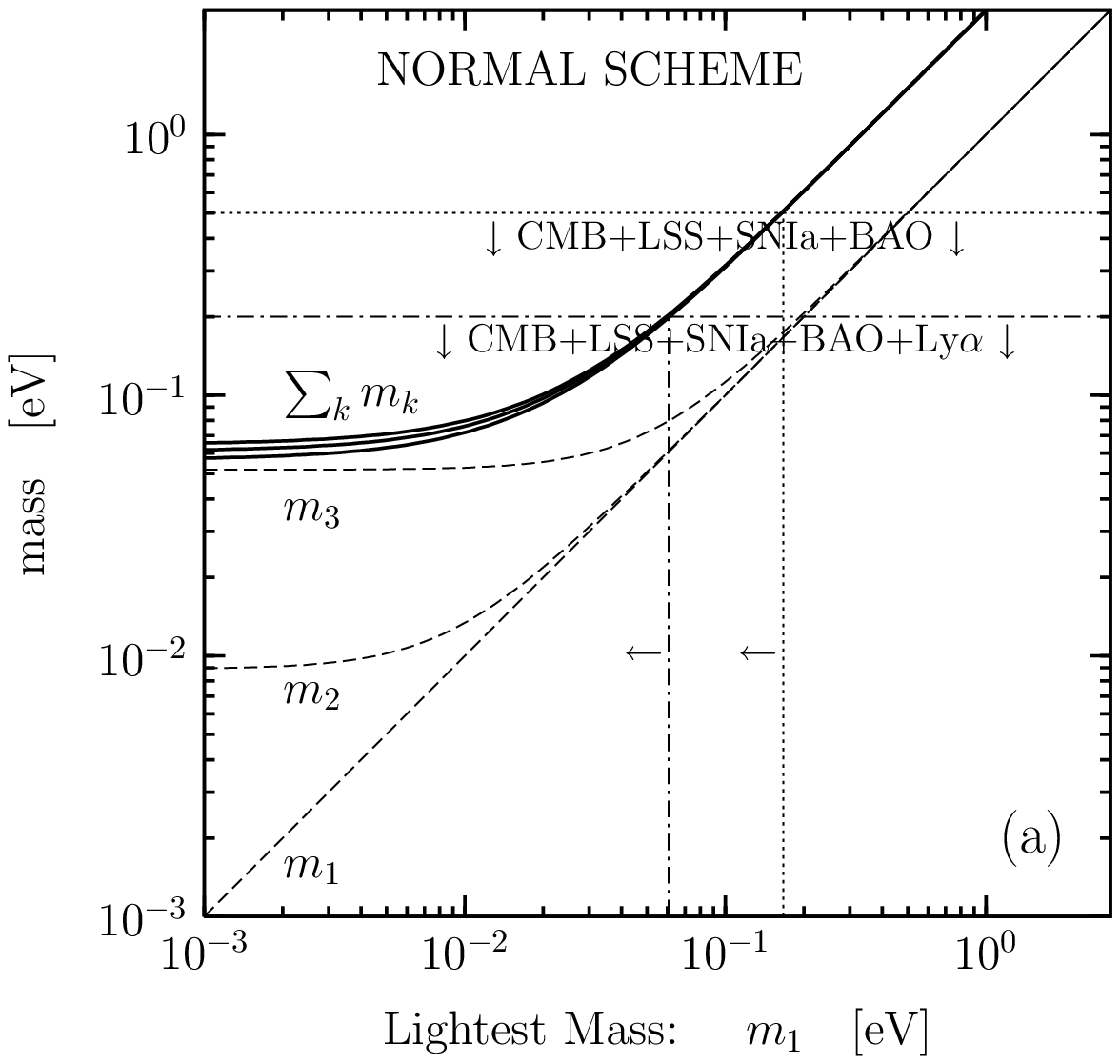}
\end{center}
\end{minipage}
\hfill
\begin{minipage}[t]{0.47\textwidth}
\begin{center}
\includegraphics*[bb=102 430 437 748, width=0.95\textwidth]{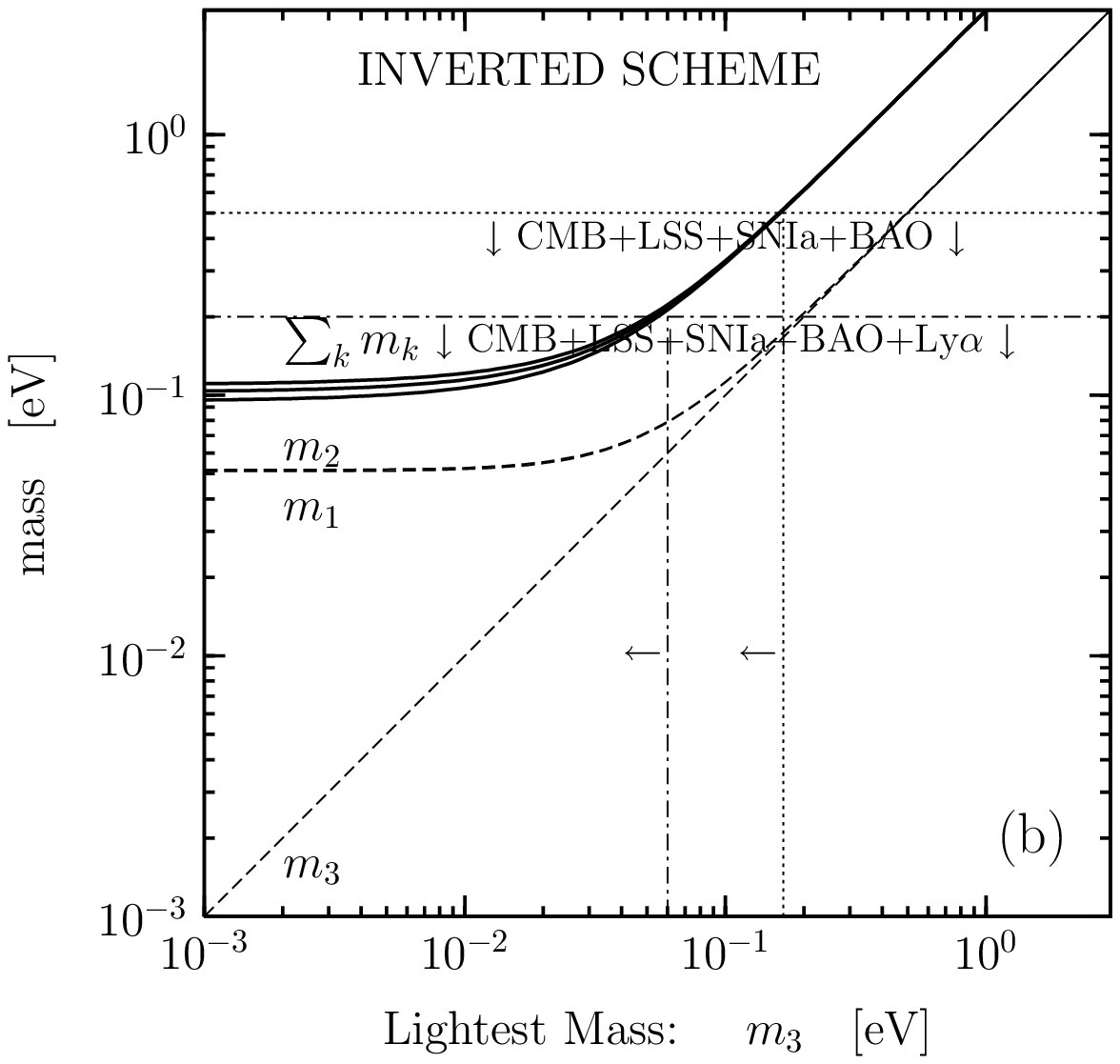}
\end{center}
\end{minipage}
\caption{ \label{cosmo}
Sum of neutrino masses in the two schemes of three-neutrino mixing
indicated by neutrino oscillation data,
as a function of the lightest mass
($m_{1}$ in the normal scheme in (a)
and
$m_{3}$ in the inverted scheme in (b)).
The three solid lines represent the best-fit and $2\sigma$ uncertainty band
obtained from the squared-mass differences in Eqs.~(\ref{SOL}) and (\ref{ATM}).
The two horizontal dotted lines represent
the approximate cosmological upper bound range in Eq.~(\ref{q143}),
and the two vertical dotted lines give the corresponding upper bound range
for the lightest mass.
The dashed curves show the three individual masses.
}
\end{figure}

\subsection{Cosmological Bounds on Neutrino Masses}
\label{n043}

If neutrinos have masses of the order of 1 eV,
they constitute a so-called ``hot dark matter'',
which suppresses the power spectrum of density fluctuations
in the early universe at ``small'' scales, of the order of
1--10 Mpc
(see Ref.~\cite{astro-ph/0603494}).
The suppression depends on
the sum of neutrino masses
$ \sum_{k} m_{k} $.

Recent high precision measurements of
density fluctuations
in the Cosmic Microwave Background
(WMAP)
and
in the Large Scale Structure distribution of galaxies
(2dFGRS, SDSS),
combined with other cosmological data,
led to stringent upper limits on
$ \sum_{k} m_{k} $,
of the order of a fraction of eV
\cite{astro-ph/0602155,astro-ph/0603449,astro-ph/0604335,hep-ph/0608060}.
The most crucial type of data are the so-called Lyman-$\alpha$ forests,
which are absorption lines in the spectra of high-redshift quasars
due to intergalactic hydrogen clouds with dimensions of the order of
1--10 Mpc.
Unfortunately, the interpretation of
Lyman-$\alpha$ data
may suffer from large systematic uncertainties.
Summarizing the different limits obtained in
Refs.~\cite{astro-ph/0602155,astro-ph/0603449,astro-ph/0604335,hep-ph/0608060},
we estimate an approximate $2\sigma$ upper bound
\begin{equation}
\sum_{k} m_{k} \lesssim 0.2 - 0.5 \, \mathrm{eV}
\,,
\label{q143}
\end{equation}
with the extremes reached with or without Lyman-$\alpha$ data.
These limits are shown in Fig.~\ref{cosmo},
where we have plotted the value of
$ \sum_{k} m_{k} $
as a function
of the unknown value of the lightest mass,
using the values of the squared-mass differences
in Eqs.~(\ref{SOL}) and (\ref{ATM}).
One can see that cosmological measurements are starting to explore
the interesting region in which
the tree neutrinos are not quasi-degenerate.
In the future, the inverted scheme can be excluded by an upper bound
of about
$9 \times 10^{-2} \, \mathrm{eV}$
on the sum of neutrino masses.

\begin{figure}[t!]
\begin{minipage}[t]{0.47\textwidth}
\begin{center}
\includegraphics*[bb=104 427 441 749, width=0.95\textwidth]{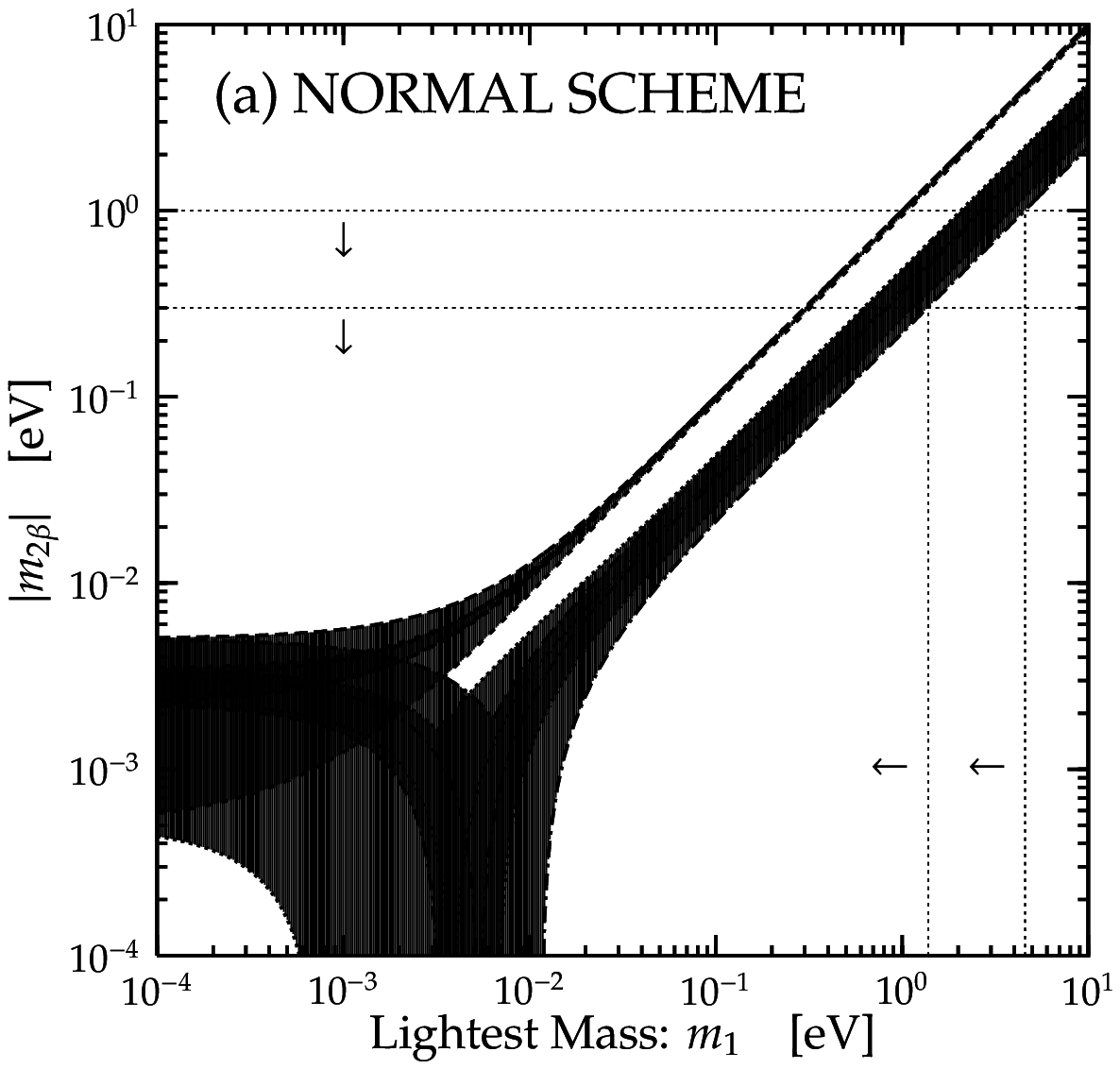}
\end{center}
\end{minipage}
\hfill
\begin{minipage}[t]{0.47\textwidth}
\begin{center}
\includegraphics*[bb=104 427 441 749, width=0.95\textwidth]{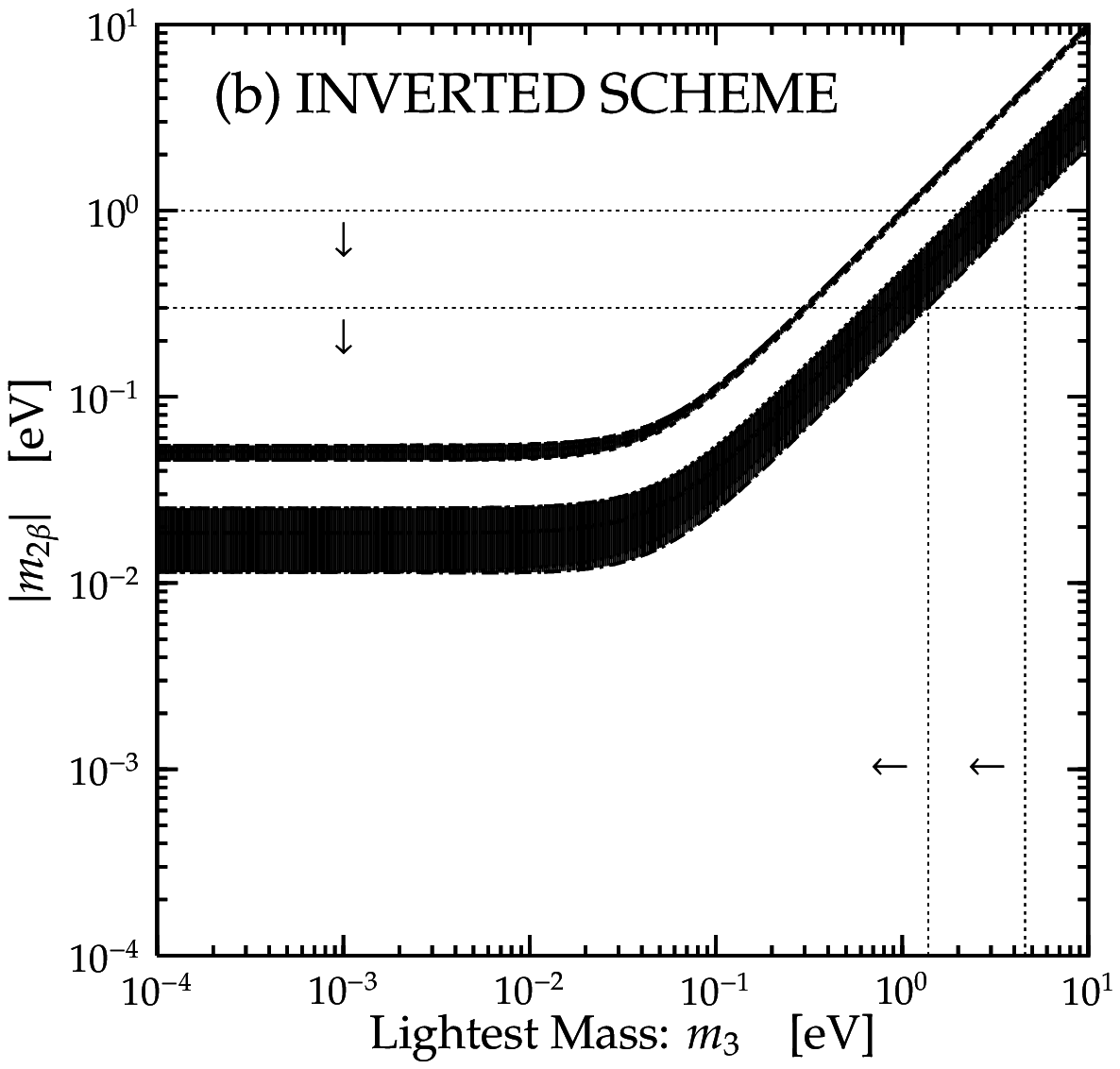}
\end{center}
\end{minipage}
\caption{ \label{db}
Absolute value $|m_{2\beta}|$ of the effective Majorana neutrino mass in
$2\beta_{0\nu}$ decay as a function
of the lightest mass $m_{1}$ in the normal scheme (a) and $m_{3}$ in the inverted scheme (b).
The white areas in the strips need CP violation.
The two horizontal dotted lines correspond to the extremes of the upper bound range
in Eq.~(\ref{d09}). The two vertical dotted lines show the corresponding
upper bounds for $m_{1}$ (a) and $m_{3}$ (b).
}
\end{figure}

\subsection{Neutrinoless Double-Beta Decay}
\label{n044}

Neutrinoless double-$\beta$ decay
($2\beta_{0\nu}$)
is a very important process,
because it is not only sensitive to the absolute value of neutrino masses,
but mainly because it is allowed only if neutrinos
are Majorana particles
\cite{Schechter:1982bd,Takasugi:1984xr}.
The observation of neutrinoless double-$\beta$ decay
would represent a discovery of a new type of particles,
Majorana particles.
This would be
a fundamental improvement in our understanding of nature.

Neutrinoless double-$\beta$ decays are processes of the type
$
\mathcal{N}(A,Z)
\to
\mathcal{N}(A,Z\pm2)
+
e^{\mp}
+
e^{\mp}
$,
in which no neutrino is emitted,
with a change of two units of the total lepton number.
These processes are forbidden in the Standard Model.
The $2\beta_{0\nu}$ half-life of a nucleus $\mathcal{N}$ is given by
(see Refs.~\cite{Doi:1985dx,Suhonen:1998ck,Civitarese:2002tu,hep-ph/0405078})
\begin{equation}
[ T_{1/2}^{0\nu}(\mathcal{N}) ]^{-1}
=
G_{0\nu}^{\mathcal{N}}
\,
|\mathcal{M}_{0\nu}^{\mathcal{N}}|^{2}
\,
\frac{ |m_{2\beta}|^{2} }{ m_{e}^{2} }
\,,
\label{n081}
\end{equation}
where $G_{0\nu}^{\mathcal{N}}$ is the phase-space factor,
$\mathcal{M}_{0\nu}^{\mathcal{N}}$
is the nuclear matrix element
and
\begin{equation}
m_{2\beta}
=
\sum_{k=1}^{3}
U_{ek}^2 \, m_{k}
\label{d03}
\end{equation}
is the effective Majorana mass.
The phase space factor can be calculated with small uncertainties
(see, for example, Table~3.4 of Ref.~\cite{Doi:1985dx} and Table~6 of Ref.~\cite{Suhonen:1998ck}).

In spite of many experimental efforts,
so far no experiment observed an unquestionable signal\footnote{
There is a claim of an observation of the $2\beta^{-}_{0\nu}$ decay of $^{76}\text{Ge}$
with
$ T_{1/2}^{0\nu}({}^{76}\mathrm{Ge}) = 1.19 {}^{+1.00}_{-0.17} \times 10^{25} \, \text{y} $
\cite{hep-ph/0201231,hep-ph/0404088}.
However, this measurement is rather controversial
\cite{hep-ph/0201291,hep-ex/0202018,hep-ex/0309016}.
The issue can only be settled by future experiments
(see Ref.~\cite{hep-ph/0405078}).
}.
The most stringent bound
has been obtained in the Heidelberg-Moscow ${}^{76}\mathrm{Ge}$ experiment \cite{Klapdor-Kleingrothaus:2001yx}:
\begin{equation}
T_{1/2}^{0\nu}({}^{76}\mathrm{Ge})
>
1.9 \times 10^{25} \, \mathrm{y}
\qquad
[90\% \, \text{CL}]
\,.
\label{d06}
\end{equation}
The IGEX experiment \cite{Aalseth:2002rf}
obtained the comparable limit
$
T_{1/2}^{0\nu}({}^{76}\mathrm{Ge})
>
1.57 \times 10^{25} \, \mathrm{y}
$
[90\% CL].
For the future,
many new $2\beta_{0\nu}$ experiments are planned and under preparation
(see Refs.~\cite{hep-ph/0405078,hep-ph/0412300}),
since the quest for the Majorana nature of neutrinos
is of fundamental importance.

The extraction of the value of $|m_{2\beta}|$
from the data
has unfortunately a large systematic uncertainty,
which is due to the
large theoretical uncertainty in the evaluation of the
nuclear matrix element
$\mathcal{M}_{0\nu}$
(see Refs.~\cite{Civitarese:2002tu,hep-ph/0405078}).
In the following, we will use as a possible range for the
nuclear matrix element
$|\mathcal{M}_{0\nu}|$
the interval which covers the results of reliable calculations
listed in Tab.~2 of Ref.~\cite{hep-ph/0405078}:
\begin{equation}
1.5
\lesssim
|\mathcal{M}_{0\nu}^{^{76}\text{Ge}}|
\lesssim
4.6
\,,
\label{d04}
\end{equation}
which corresponds to an uncertainty of a factor of 3 for the
determination of $|m_{2\beta}|$ from $T_{1/2}^{0\nu}({}^{76}\mathrm{Ge})$.
Using the range (\ref{d04}),
the upper bound (\ref{d06}) implies
($
G_{0\nu}^{^{76}\text{Ge}}
=
6.31 \times 10^{-15} \, \text{y}^{-1}
$)
\begin{equation}
|m_{2\beta}|
\lesssim
0.3 - 1.0 \, \mathrm{eV}
\,.
\label{d09}
\end{equation}
% In the standard parameterization of the mixing matrix,
% the effective Majorana mass is given by
% \begin{equation}
% m_{2\beta}
% =
% c_{12}^2 \, c_{13}^2 \, m_{1}
% +
% s_{12}^2 \, c_{13}^2 \, e^{i\alpha_{21}} \, m_{2}
% +
% s_{13}^2 \, e^{i\alpha_{31}} \, m_3
% \,,
% \label{d003}
% \end{equation}
% where
% $\alpha_{21}$ and $\alpha_{31}$
% are Majorana phases
% (see Refs.~\cite{Bilenky:1987ty,hep-ph/9812360,hep-ph/0211462,hep-ph/0310238}),
% whose values are unknown.

Figure~\ref{db} shows the allowed range for
$|m_{2\beta}|$
as a function
of the unknown value of the lightest mass,
using the values of the oscillation parameters
in Eqs.~(\ref{SOL}), (\ref{ATM}), (\ref{m140}), (\ref{151}) and (\ref{152}).
One can see that, in the region where the lightest mass is very small,
the allowed ranges for $|m_{2\beta}|$
in the normal and inverted schemes are dramatically different.
This is due to the fact that
in the normal scheme strong cancellations between the contributions
of $m_{2}$ and $m_3$ are possible,
whereas in the inverted scheme
the contributions of $m_{1}$ and $m_{2}$ cannot cancel, because maximal mixing
in the 1$-$2 sector is excluded by solar data
($\vartheta_{12}<\pi/4$ at $5.8\sigma$ \cite{Bahcall:2004ut}).
On the other hand,
there is no difference between the normal and inverted schemes
in the quasi-degenerate region,
which is probed by the present data.
From Fig.~\ref{db} one can see that, in the future,
the normal and inverted schemes may be distinguished by reaching a sensitivity
of about $10^{-2} \, \mathrm{eV}$.

\section{Conclusions}
\label{n045}

The results of neutrino oscillation experiments
have shown that neutrinos are massive particles,
there is a hierarchy of squared mass differences
and the mixing matrix is bilarge
i.e. with two large and one small mixing angles.

From the theoretical point of view, it is very likely that
massive neutrinos are Majorana particles,
with a small mass connected to new high-energy physics beyond the Standard Model
by a see-saw type relation.
An intense experimental effort is under way in the search for
neutrinoless double-$\beta$ decay,
which is the most accessible signal of the Majorana nature of massive neutrinos.

Since neutrino oscillations depend on the differences of the squared neutrino masses,
the absolute scale of neutrino masses is still not known,
except for upper bounds obtained in
$\beta$ decay and neutrinoless double-$\beta$ decay experiments
and through cosmological observations.

The measurement of the effective electron neutrino mass in $\beta$ decay
experiments is robust but very difficult.
The future KATRIN experiment \cite{hep-ex/0309007} will reach a sensitivity of about $0.2 \, \text{eV}$.

Cosmological observations have already pushed the upper limit for the sum of the neutrino masses
at a few tenths of eV,
in the interesting region in which
the tree neutrinos are not quasi-degenerate.
Significant improvements are expected in the near future
(see Ref.~\cite{astro-ph/0603494}),
with the caveat that cosmological information on fundamental physical quantities
depend on the assumption of a cosmological model and on the interpretation of
astrophysical observations.

Let us finally mention that we have not considered the
indication of $\bar\nu_{\mu}\to\bar\nu_{e}$ transitions,
found in the LSND experiment
\cite{hep-ex/0104049}.
This signal is under investigation
in the MiniBooNE experiment at Fermilab
\cite{hep-ex/0602018}.
This check is important,
because a confirmation of the LSND signal could require an extension
of the three-neutrino mixing scheme
(see Refs.~\cite{hep-ph/9812360,hep-ph/0202058}).

\end{document}